# High-Entropy Enhanced Negative Thermal Expansion Perfomance in Antiperovkites

Xiuliang Yuan[a], Bing Wang[a], Ying Sun[a,*], Huaiming Guo[a], Kewen Shi[b], Sihao Deng[c,d,†], Lunhua He[c,e,f], Huiqing Lu[g], Hong Zhang[h,i,‡], Shengdi Xu[a], Yi Du[a], Weichang Hao[a], Shengqi Chu[d], Zhijie Ma[a], Shihai An[a], Jin Cui[b], Dongmei Hu[a], Huiming Han[a], Cong Wang[b,§]

a   School of Physics, Beihang University, Beijing 100191, China
b   School of Integrated Circuit Science and Engineering, Beihang University, Beijing 100191, China
c   Spallation Neutron Source Science Center, Dongguan 523803, China
d   Institute of High Energy Physics, Chinese Academy of Sciences, Beijing 100049, China
e   Beijing National Laboratory for Condensed Matter Physics, Institute of Physics, Chinese Academic of Sciences, Beijing 100190, China.
f   Songshan Lake Materials Laboratory, Dongguan, Guangdong 523808, China
g   Key Laboratory for Photonic and Electronic Bandgap Materials, Ministry of Education, School of Physics and Electronic Engineering, Harbin Normal University, Harbin 150025, China
h   School of Materials and Energy, Electron Microscopy Centre, Yunnan University, Kunming 650500, P.R. China
i   School of Materials and Energy, Electron Microscopy Centre, Lanzhou University, Lanzhou 730000, P. R. China

# 1   Abstract

The negative thermal expansion (NTE) materials, which can act as thermal-expansion compensators to counteract the positive thermal expansion, have great applications merit in precision engineering. However, the exploration of NTE behavior with a wide temperature range has reached its upper ceiling through traditional doping strategies due to composition limitations. The unique sluggish characteristic in phase transition and extended optimization space in recent high entropy systems has great potential to broaden the temperature range in electronic transitions-induced NTE materials. Mn-based anti-perovskites offer an ideal platform for the exploration of high entropy NTE material due to their abundant element selection and controllable NTE performance. In this paper, the high entropy strategy is first introduced to broaden the NTE temperature range by relaxing the abrupt phase transition in Mn-based anti-perovskite nitride. We propose an empirical screening method to synthesize the high-entropy anti-perovskite (HEAP). By means of neutron diffraction analysis of the typical HEAP $Mn_3Cu_{0.2}Zn_{0.2}Ga_{0.2}Ge_{0.2}Mn_{0.2}N$, it is found that magnetic phase separation from anti-ferromagnetic CII to paramagnetic CI surviving in an ultra-wide temperature range of 5 K $\leq T \leq$ 350 K ($\Delta T$ = 345 K), revealing a unique sluggish characteristic. Consequently, a remarkable NTE behavior (up to $\Delta T$ = 235 K, 5 K $\leq T \leq$ 240 K) with a coefficient of thermal expansion (CTE) of $-4.7 \times 10^{-6}$ /K, has been obtained in HEAP. It is worth noting that the temperature range is two or three times wider than that of traditional low-entropy doping systems. The sluggish characteristic has been further experimentally proved to come from disturbed phase transition dynamics due to distortion in atomic spacing and chemical environmental fluctuation observed by the spherical aberration-corrected electron microscope. Our demonstration provides a unique paradigm for broadening the temperature range of NTE materials induced by phase transition through entropy engineering.

*  E-mail address: sunying@buaa.edu.cn
†  E-mail address: dengsh@ihep.ac.cn
‡  E-mail address: hongzhang@lzu.edu.cn
§  E-mail address: congwang@buaa.edu.cn

# 2 Introduction

The controllable thermal expansion is of great significance for improving structural reliability and safety in engineering applications such as spacecraft(*1*), precision space-based telescopes(*2*), and containers for liquid hydrogen or liquefied natural gas[3]. The negative thermal expansion (NTE) material with the characteristic of volume contraction when heated can adjust the overall coefficient of thermal expansion (CTE) of materials, making it useful as a thermal expansion compensator[4-6]. In general, a practical NTE material should have a wide working temperature range to accommodate the significant temperature fluctuations in actual environment temperature[7].

In the past two decades, numerous NTE materials have been discovered in various systems[8, 9]. Among these, NTE from electronic transitions, particularly the first-order transition accompanied by an abrupt volume change, is observed in anti-perovskite manganese nitrides, La(Fe, Si)$_{13}$, LaCu$_3$Fe$_4$O$_{12}$, YbInCu$_4$, and SmS. These materials exhibit a maximal NTE magnitude[9]. Relaxing this abrupt phase transition to achieve a broad temperature range of NTE behavior is an important research field. However, it has reached the ceiling of performance optimization through chemical element doping over the past decade due to composition limitations[5, 6, 10, 11]. The recently proposed high-entropy strategy offers an extended and optimized composition space, breaking through the traditional approach of chemical element doping[12, 13]. More importantly, the unique sluggish characteristic of phase transition in high-entropy materials, caused by high lattice distortion, would result in a more effective relaxation of abrupt phase transition[14-16], making it great potential to broaden the temperature range in electronic transitions induced NTE materials. However, the high-entropy strategy used to broaden the temperature range of NTE has not yet been explored.

As a typical NTE material, Mn-based anti-perovskite nitrides Mn$_3A$N, where $A$ is usually a transition metal, metalloid, or rare earth metal, possess a colossal magnitude of NTE due to magnetic order-driven volume contraction[9]. This makes it an ideal candidate for studying the high-entropy effects in broadening the NTE temperature range. The high-entropy anti-perovskites (Mn-based nitrides) can be defined as a single phase with high configurational entropy (greater than 1.5R, where R is the universal gas constant), which originates from the disorderly occupation of five or more types of atoms at the same sublattice site(*15, 17-23*). In the typical cubic anti-perovskite structure, the 1*a* site (located at the corner of the cubic) is a more suitable location for achieving the desired disorder occupation due to its richer element selection in performance optimization compared to 1*b* (body center) and 3*c* (face center) sites(*6, 7, 10, 11, 24-29*). However, chemical segregation or low-dimensional phase decomposition at the grain boundaries, driven by the minimization of the system's free energy, often hinders the formation of single-phase high-entropy materials(*30, 31*). At this point, the design of components is crucial and full of challenges for stabilizing the single-phase high-entropy anti-perovskites (HEAPs).

In this work, we propose an empirical screening method based on experimental results to screen for the potential presence of the HEAP. Based on this method, a total of 37 HEAPs have been proposed. More importantly, in typical HEAP $Mn_3Cu_{0.2}Zn_{0.2}Ga_{0.2}Ge_{0.2}Mn_{0.2}N$, the unique sluggish characteristic in the form of ultra-wide temperature range of magnetic phase transition process due to the distortion significantly broadens the NTE temperature range with temperature span of $\Delta T$ = 235 K (5 K $\leq T \leq$ 240 K), which is much wider than that of traditional low-entropy doping systems(*6, 32*). It is worth noting that this unique paradigm of entropy engineering used in the broadening NTE temperature range through sluggish characteristics is proposed for the first time and will open an avenue to the design of NTE or ZTE materials.

## 3  Results and Discussion

Firstly, we developed an effective strategy to obtain HEAPs experimentally by overcoming the synthesis difficulties. In the past two decades, more than one hundred Mn-based anti-perovskite nitrides, including 97 compounds doped at 1*a* site, has been discovered. In these doped compounds, 1*a* site is shared by different atoms with disordered occupation, which are involved with 24 different types of elements. We summarize the solubility of the two related elements in an anti-perovskite structure using a checkerboard diagram (see Figure 1(b)). From the checkerboard diagram, certain atoms such as Cu, Zn, Ga, Ge, and Sn (or Mn) are surrounded by a red (or blue) dotted line, indicating their mutual solubility with each other. Based on the mutual solubility of these five atoms, we make a radical inference that they have the potential to crystallize in an anti-perovskite structure with Mn and N atoms. This can be illustrated in Figure 1(a), where the five elements associated with ten binary-dopant compounds can combine to form a single quantity-dopant compound.

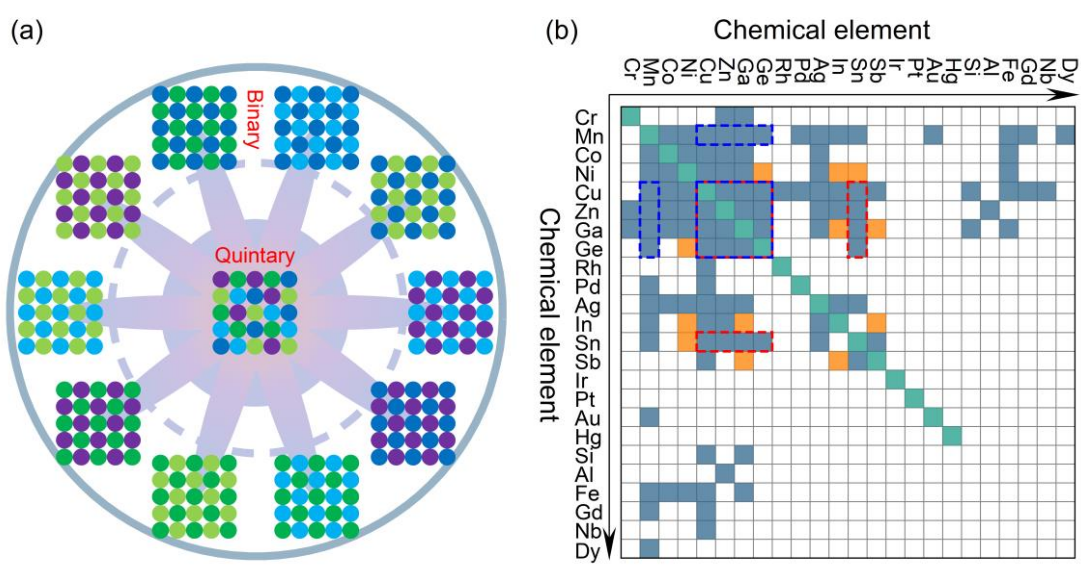

Figure 1(a) shows that the existence of a single quinary-dopant compound can be predicted by the five binary-dopant compounds. (b) The solubility checkerboard diagram of the two elements related to the anti-perovskite structure. The dark blue grid represents the solid solution formed by the two related elements, Mn and N, in the anti-perovskite structure. The orange grid does not form a solid solution. The white grid represents the unknown solubility. The cyan grid represents the undoped compound.

Based on the empirical screening method, the existence of up to 37 additional HEAPs is predicted. The crystal structure of synthesized samples is checked by room temperature X-ray diffraction (XRD) (see Figure S1 and Table

S1, Supplementary information). The Bragg peaks shown in the XRD pattern can be indexed to a cubic anti-perovskite structure (space group P$m$-3$m$) (see the insert of Figure 2(a)). We conducted a Rietveld refinement on the XRD pattern of Mn$_3$Cu$_{0.2}$Zn$_{0.2}$Ga$_{0.2}$Ge$_{0.2}$Mn$_{0.2}$N at 410K using the anti-perovskite phase (Mn at 3$c$ (0.5, 0.5, 0), Cu/Zn/Ga/Ge/Mn (HE) at 1$a$ (0, 0, 0), and N at 1$b$ (0.5, 0.5, 0.5)). Additionally, there is a trace amount of MnO impurity. The results show a good agreement between the calculated pattern and the observed pattern, with a lattice constant of 3.90044 (6) Å for HEAP (see Figure 2(a)).

The configurational entropy of anti-perovskite can be given by(33, 34):

$$S_{\text{config}} = -R \left[ 3 \left( \sum_{i=1}^{\text{Mn}} x_i \ln x_i \right)_{3c-\text{site}} + \left( \sum_{j=1}^{M} x_j \ln x_j \right)_{1a-\text{site}} + \left( \sum_{k=1}^{X} x_k \ln x_k \right)_{1b-\text{site}} \right] \quad (1)$$

where $x_{i(j,k)}$ represent the mole fraction of the element $i(j,k)$ located at site 3$c$ (1$a$, 1$b$). The calculated configurational entropy is the molar configurational entropy. The configurational entropy of the five-element co-doped anti-perovskite manganese nitride Mn$_3$Cu$_{0.2}$Zn$_{0.2}$Ga$_{0.2}$Ge$_{0.2}$Mn$_{0.2}$N is equal to 1.61R, which is higher than the lower limit of the configurational entropy value (1.5R) for a high entropy system. Therefore, the high-entropy anti-perovskite could be precisely defined and confirmed experimentally.

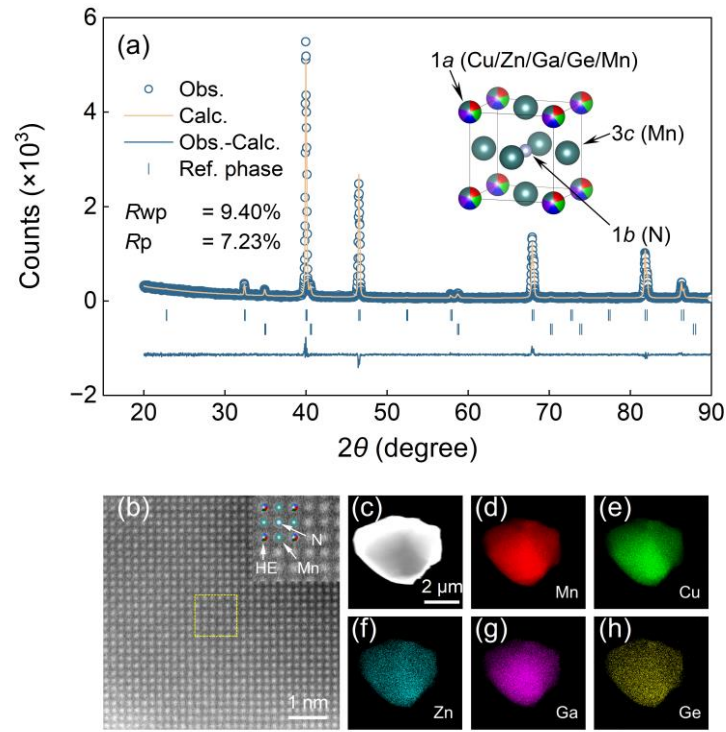

Figure 2(a) The result of the Rietveld refinement of XRD patterns of Mn$_3$Cu$_{0.2}$Zn$_{0.2}$Ga$_{0.2}$Ge$_{0.2}$Mn$_{0.2}$N at 410K. The figure illustrates the crystal structure of high-entropy perovskites. (b) Spherical aberration-corrected electron microscope of HEAP along [001]. The top right illustration gives the enlarged image with an atomic mask. (c) High-angle annular dark-field (HAADF) image on the selected grains. The elemental area profiles for (d) Mn, (e) Cu, (f) Zn, (g) Ga, and (h) Ge. Each of them appears homogenous on the selected grains.

The results of the electron microscope provide a detailed analysis of the microstructure and distribution of chemical elements. Figure 2(b) shows the atomic resolution HAADF image along the axis [001]. The mask of disordered HE and ordered Mn and N of HEAP structure is shown in the enlarged illustration of Figure 2(b). The elements Mn, Cu, Zn, Ga, and Ge in Figure 2(d), (e), (f), (g), and (h) exhibit nearly identical spatial distributions, indicating a disordered atomic arrangement of Mn, Cu, Zn, Ga, and Ge. This indicates that these five atoms collectively crystallize in an anti-perovskite structure with Mn and N atoms.

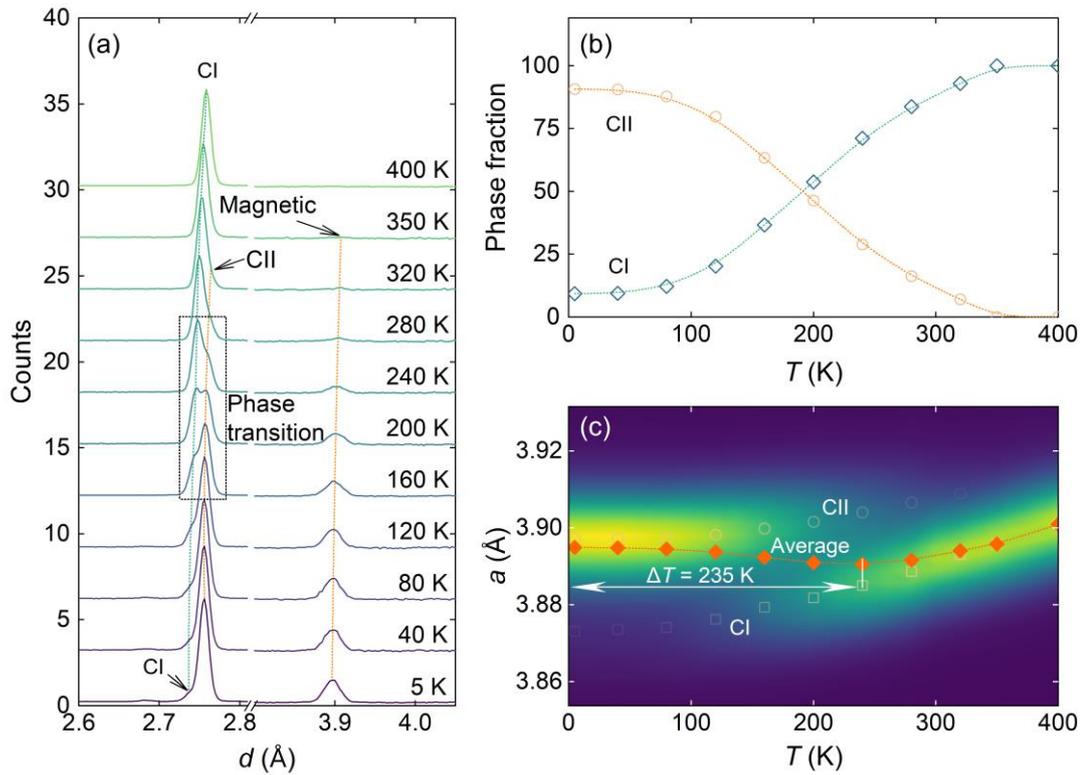

Figure 3 (a) Temperature-dependent NPD patterns for HEAP. (b) Temperature-dependent phase fraction of cubic phases C1 and C2. (c) Temperature-dependent lattice constant of cubic phase CI, CII, and the average lattice constant.

Now, we turn to discuss the NTE behavior of the high-entropy anti-perovskite manganese nitride. Firstly, we investigate the phase separation phenomenon observed in room temperature X-ray diffraction because it is a typical feature of NTE behavior in $Mn_3AN$ (see Figure S2, Supplementary information). In these systems, NTE behavior is derived from the phase transition from a bigger cubic phase to a smaller cubic phase at a certain temperature range. As seen in Figure 3(a), the sharp phase transition process is indeed detected at 160 K ≤ T ≤ 280 K in the temperature-dependent neutron powder diffraction (NPD) patterns. The temperature-dependent phase fraction is quantized, as shown in Figure 3(b). Notably, the phase separation phenomenon extends to an ultra-wide temperature range of 5 K ≤ $T$ ≤ 350 K, which covers room temperature. The two separated cubic phases possess different lattice constants sharing the same space group of $Pm$-$3m$. In Figure 3(c), both of them display positive thermal expansion behavior, but phase CII (with a bigger unit cell) gradually transforms into phase CI (with a smaller unit cell). We emphasize here that phase separation survival in such a wide temperature range is not a common phenomenon in low entropy anti-perovskite. For instance, the temperature range of phase separation in $Mn_3Cu_{0.5}Ge_{0.5}N$ and $Mn_3Ga_{0.3}Sn_{0.7}N$ are 280 K $\leq T \leq$ 365 K and 418 K $\leq T \leq$ 454 K, respectively. The considerable difference between HEAP and low entropy anti-perovskites suggests a distinct mechanism for the ultra-wide temperature range of the phase separation phenomenon.

Secondly, the weighted average lattice constant, taking into account the temperature-dependent phase fraction of CII and CI, exhibits a remarkable NTE behavior at the temperature range of 5 K $\leq T \leq$ 240 K ($\Delta T$ = 235 K) with an average coefficient of thermal expansion (CTE) value of $-4.7\times10^{-6}$ /K. Although the working temperature range of NTE behavior is not extended to the entire temperature range of phase separation, it is still significantly wider than that of the traditional low-entropy doping system ($\Delta T \sim$ 122 K for $Mn_3Ga_{0.7}Ge_{0.3}N_{0.88}C_{0.12}$(6), $\Delta T \sim$ 60 K for $Mn_3Ga_{0.6}Ge_{0.4}N$(32)).

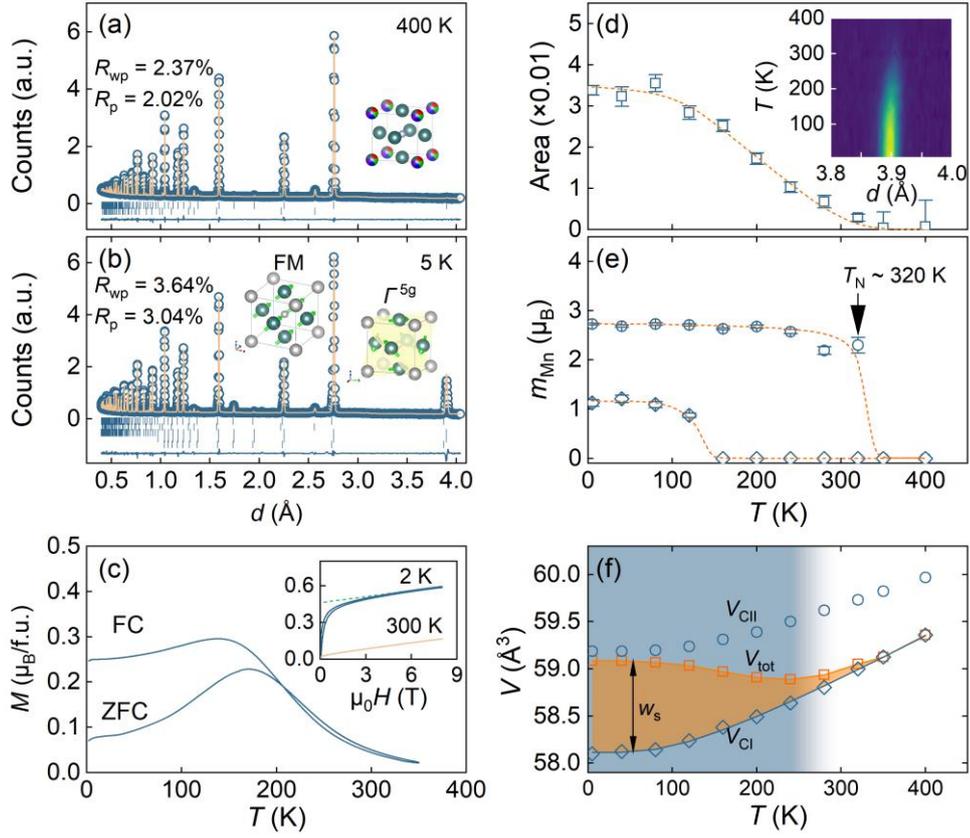

Figure 4(a) Rietveld refinement results of neutron diffraction patterns at 400 K. The insert shows the crystal structure of HEAP. (b) Nuclear and $\Gamma^{5g}$ with ferromagnetic components at 5 K. The circles represent the experimental intensities ($I_{obs}$), the upper solid line represents the calculated intensities ($I_{calc}$), and the lower solid line represents the difference between observed and calculated intensities ($I_{obs} - I_{calc}$). The vertical bars mark the angular positions of the nuclear Bragg peaks of the cubic phase CI (the first row), the nuclear Bragg peaks of MnO (the second row), the nuclear Bragg peaks of the cubic phase CII (the third row), and the magnetic Bragg peaks of noncollinear triangular antiferromagnetism $\Gamma^{5g}$ (the fourth row). (c) The temperature-dependent magnetization, including the zero field cooling (ZFC) curve and the field cooling (FC) curve. The insert shows the isothermal magnetization curve at 2 K and 300 K. (d) The temperature-dependent area of the Bragg peak for the characteristic peak of the $\Gamma^{5g}$ magnetic structure. The insert shows the contour diagram of the characteristic peak of the $\Gamma^{5g}$ magnetic structure. (e) The temperature-dependent Mn moment of the $\Gamma^{5g}$ magnetic structure and the ferromagnetic component. (f) The temperature-dependent volume of the unit cell of phases CI and CII. The solid line represents the predicted result of the Debye equation considering the lattice vibrations.

The temperature-dependent NPD patterns also give the evolution of magnetic configuration. We have conducted the magnetic structure analysis on NPD patterns at all temperatures. Figure 4(b) and (c) gives the typical results of Rietveld refinement for the magnetic disordering state at 400 K and ordering state at 5 K. The cubic anti-perovskite structure at 400 K analyzed by NPD is consistent with X-ray diffraction in Figure 2(a). At 5 K, the spin of Mn at $3c$ site forms a so-called $\Gamma^{5g}$ antiferromagnetic structure with a ferromagnetic component ~ 1 $\mu_B$ (see the insert in Figure 4(b)). This ferromagnetic component disappears when the temperature increases to 150 K, which is consistent with macroscopic magnetic measurement results in Figure 4(c) and Figure 4(e). On the other hand, as seen in Figure 4(d) and Figure 4(e), the antiferromagnetic component disappears completely at 350 K, higher than the ferromagnetic component. It is worth emphasizing that the phase fraction of the magnetic structure of $\Gamma^{5g}$ with ferromagnetic component decreases gradually with the increase of temperature, while the moment of Mn remains unchanged (see Figure 4(e)). The NTE behavior exists in the temperature range of dramatic change for magnetic phase fraction. In other words, The NTE in HEAP occurs as a result of the volume contraction accompanied by magnetic ordering, the so-called magneto volume effect (MVE)(*27*). In a low entropy system, the MVE is quantified by spontaneous volume magnetostriction $w_s$, which is linear- or square-dependent of magnetic moment dependency(*3, 35-37*). However, this inference is not suitable for the present high entropy anti-perovskite due to the disagreement in temperature-independent magnetic moment (~ 2.7 $\mu_B$ / Mn) and temperature-dependent magnetostriction $w_s$ (see Figure 4(f)), suggesting a more complicated mechanism of magnetovolume effect. We speculate the unique sluggish characteristic in high entropy systems plays an important role in MVE-induced NTE behavior.

The sluggish characteristic in phase transition is an inherent and universal feature in high entropy systems due to lattice distortion. For instance, the sluggish transformation dynamics can be observed in systems with phase transformation, such as CoCrFeNiAl high entropy alloy and TiZrHfCuNiCo high-entropy intermetallics(*14, 16*). For the current HEAP, the ultra-wide temperature range of the phase separation phenomenon reveals a remarkable enhancement of sluggish characteristics compared to low entropy systems. Further, inherent lattice distortion can be directly observed and quantized by a spherical aberration-corrected electron microscope (will be discussed later). The consequence induced by lattice distortion in magnetism is illustrated here. As seen in the insert of Figure 4(c), the isothermal magnetization curve at 2 K exhibits a ferromagnetic feature, which is counterintuitive in a long-range magnetic ordered antiferromagnet. Careful analysis of temperature-dependent NPD indicates the presence of another ferromagnetic ordering. Actually, the presence of ferromagnetic features in high entropy systems is not rare. For example, in La(Co$_{0.2}$Cr$_{0.2}$Fe$_{0.2}$Mn$_{0.2}$Ni$_{0.2}$)O$_3$, the presence of a ferromagnetic feature can be observed through Mössbauer spectra(*38*). These ferromagnetic features are believed to originate from a combination of canted antiferromagnetic arrangement, locally uncompensated spins, small ferromagnetic clusters, or both(*38*). For the current HEAP, the ferromagnetic feature mainly arises from the uncompensated spins and ferromagnetic clusters induced by the magnetic Mn atom at the 3c site, as well as the canted antiferromagnetic arrangement resulting from local distortion caused by the highly disordered arrangement of atoms with different atomic radii.

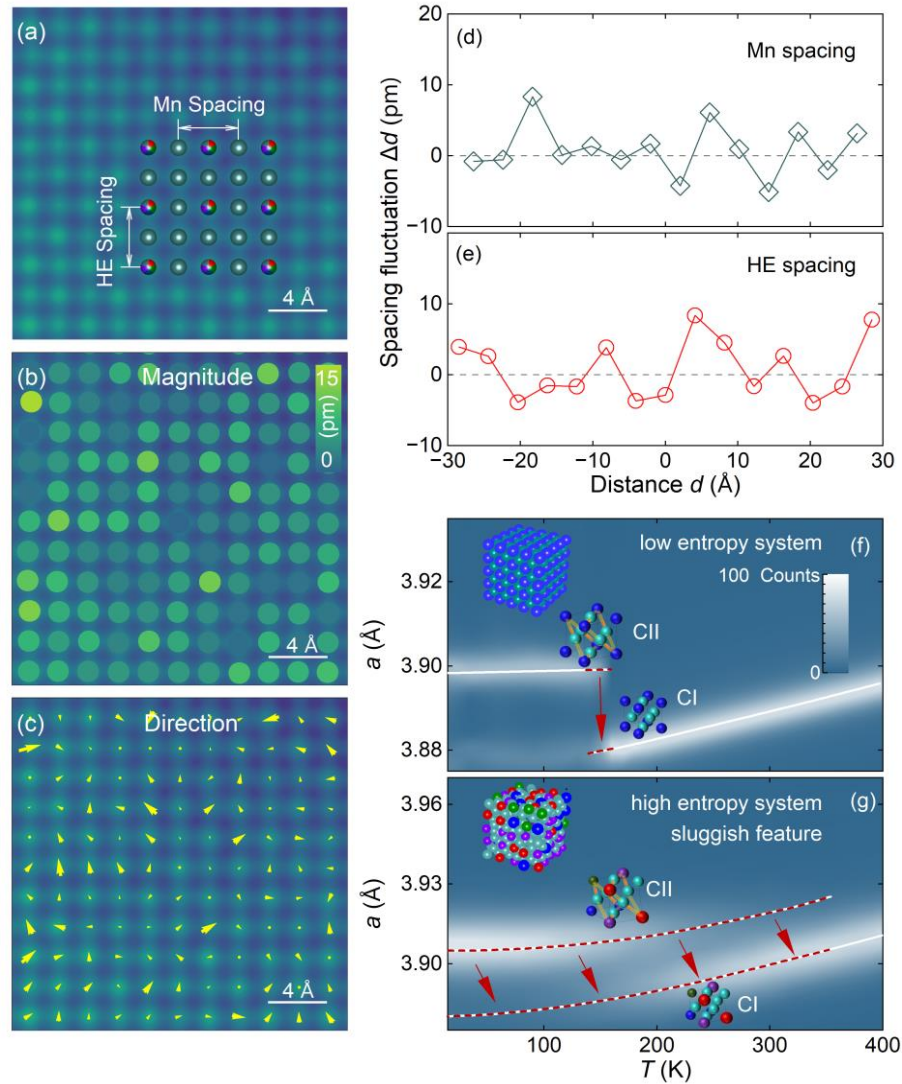

Figure 5 (a)the fitted STEM image by superposition of Gaussian peaks. The musk of crystal structure shows the atomic type of atom columns. (b) the lattice distortion illustrated by the mappings of magnitude and (c) direction of displacements for the atom column position. The spacing fluctuation for (d) Mn atoms (at $3c$ site) and (e) disordered atoms (at $1a$ site). (f) The magnetic phase transition (red arrow) at a narrow temperature range (red dotted line) for low entropy systems. (g) The magnetic phase transition (red arrow) at an ultra-wide temperature range (red dotted line) for HEAP. The bright areas shown in the background of (f) and (g) denote the diffraction intensity of the magnetic phase ($\varGamma^{5g}$, up) and paramagnetic phase (down).

Then, the direct observation of lattice distortion (quantization in atom displacement and spacing fluctuation) in the high entropy anti-perovskite Mn$_3$Cu$_{0.2}$Zn$_{0.2}$Ga$_{0.2}$Ge$_{0.2}$Mn$_{0.2}$N is shown in Figure 5 using a high-resolution scanning transmission electron microscopy (STEM) image. For Mn-based anti-perovskite structures, the atoms should be located at high-symmetry lattice sites, such as the 1$a$ and 3$c$ sites. However, in HEAP, the deviation from these high-symmetry lattice sites cannot be ignored. We present the distribution of the magnitude and direction of displacement in Figure 5(b) and (c), respectively. The magnitude exhibits a random distribution ranging from 0 to 45 pm, with an average value of 9.3(5) pm. This average value is significantly greater than the systematic error of 3.6(5) pm (see Figure S6, supplementary information). The direction shows a disordered distribution, revealing a non-periodic lattice distortion(*39, 40*). In Figure 5(d) and (e), the atom spacing of Mn (at 3c site) and HE (at 1a site) exhibit random fluctuations (around 10 pm). The mean spacing is consistent with the distribution of the displacement magnitude and direction shown in Figure 5(b) and (c). The fluctuation in atom spacing can also be revealed by the significantly larger error bar in interatomic distance of HEAP than that of Mn$_3$CuN (reference) via analysis of extended X-ray absorption fine structure (see Figure S5 and Table S2, supplementary information).

Now we turn to discuss the high entropy effect on the broadening of temperature range of NTE behavior. The NTE behavior of magnetic materials is controlled by the magnetic ordered process of specific magnetic structures. Although the mechanism of MVE for specific magnetic structures is controversial(*41-43*), the universality of magnetic ordered process controlled thermal expansion is robust: the evolution of magnetic ordering has the same tendency with the net magnetic contribution of thermal expansion $w_s$(*37, 44*). In Mn$_3$Ga$_{0.3}$Sn$_{0.7}$N(*37*) and Ho$_2$Fe$_{16}$Co(*44*), the net magnetic contribution thermal expansion is proportional to the square of the ordering parameters(*37, 44*). Therefore, relaxing the variation of ordering parameter with temperature is the key to broadening the NTE behavior. In low entropy systems, the fluctuation of interaction strength at different positions is relatively small and brings out a little fluctuation in internal energy reduction due to spin ordering (the increase of ordering parameters) at different positions. A slight drop in temperature will make the internal energy reduction (due to the change in interaction) completely compensate for the energy increase (due to entropy reduction). The related spin correlation length grows rapidly from zero to infinity at the epsilon neighborhood of critical temperature, which is the typical physical image of a continuous magnetic phase transition. Therefore, the significant change in ordering parameter occurs in a relatively narrow temperature range, which doesn't lead to a wide NTE temperature range in the low entropy system (see Figure 5(f)).

For high entropy systems, the above description is not applicable. According to Ruderman–Kittel–Kasuya–Yosida (RKKY) interaction theory(*45, 46*), the fluctuation of magnetic atom distance significantly varies the interaction strength and the magnetic phase transition temperatures. The fluctuation of magnetic atom distance for Mn – Mn atoms in HEAP will bring out the large internal stress fluctuation (~6 GPa)(*47*) and considerable shift of transition temperature (~390 K)(*48*). Although this rough estimate of transition temperature is the extrapolated result under a uniform tension situation, the large fluctuation in interaction strength reflected by transition temperature is conceivable in the high entropy system. The large fluctuation in interaction strength makes energy competition between the internal energy and entropy occur at a broadening temperature range. The relevant growth of spin correlation length is considered to be hindered or even blocked due to the large fluctuation in interaction strength and occurs at a broadening temperature range rather than the epsilon neighborhood of critical temperature. As a consequence, the variation in ordering parameters accompanied by NTE behavior occurs at a broadened temperature range (see Figure 5(g)).

# 4   Conclusion

In conclusion, the unique sluggish characteristic of phase transition is first utilized to broaden the working temperature range of negative thermal expansion (NTE) through the high entropy strategy. For this purpose, we have proposed an effective empirical screening method and screened 37 high-entropy anti-perovskites. A remarkable NTE behavior (up to $\Delta T = 235$ K, $5$ K $\leq T \leq 240$ K) with the average coefficient of thermal expansion (CTE) of $-4.7 \times 10^{-6}$ /K has been obtained in typical HEAP $Mn_3Cu_{0.2}Zn_{0.2}Ga_{0.2}Ge_{0.2}Mn_{0.2}N$, whose work temperature range is far wider than that of a conventional low-entropy doping system. The broadened NTE temperature range is ascribed to the ultra-wide temperature range ($5$ K $\leq T \leq 350$ K) of the phase separation phenomenon caused by the unique sluggish characteristic in HEAP due to intrinsic lattice distortion, which hinder the growth of spin correlation length. This work proposes a novel mechanism used for the broadening of the NTE temperature range through entropy engineering.

# 5 Methods

**Synthesis methods.** The high-entropy anti-perovskites are synthesized through a solid-state reaction using the self-made precursor $Mn_2N$ and industrial metallic powder or bulk materials with a purity of 4N. Firstly, the precursor $Mn_2N$ is synthesized through a gas-solid reaction. The industrial metallic powder with a purity of 4N is placed into a ceramic vessel and forms a layer of Mn powder (around 3mm thick) with a flat surface. The ceramic vessel is then heated to 750℃ and maintained at that temperature for 48 hours under a nitrogen flow (4N purity) in a tube furnace to facilitate the gas-solid reaction. After cooling down to room temperature, the bulk material is obtained. The remaining bulk material is ground into a powder, serving as the precursor for Mn2N. Secondly, if the final HEAP contains Ga, the $Mn_2N$ powder and the fine Ga bulk should be mixed with a molar ratio of 1.5:0.2. The mixture is placed into a ceramic vessel once again and annealed at 500℃ under a flow of argon gas for 5 hours in the same tube furnace. After cooling down to room temperature, the Ga bulk will permeate the surrounding $Mn_2N$ powders. This mixture is further combined with the remaining metallic powders in a molar fraction of 0.2. Therefore, the molar ratio between the five metallic elements is 1:1:1:1:1. Thirdly, the mixture of Mn2N and the five metallic powders is annealed at 800℃ to facilitate the solid-state reaction. The final HEAP will be obtained after cooling down to room temperature.

**X-ray diffraction measurements.** Phase identification at room temperature is conducted by X-ray diffraction (XRD) using a Cu Target (acceleration voltage 40 kV, current 40 mA) and a compact cradle sample stage on an X-ray diffractometer (D8 Advance, BRUKER AXS, Germany). The high-resolution XRD is conducted using the same X-ray diffractometer (D8 Advance, BRUKER AXS, Germany), but it is equipped with a Ge004 monochromator and a non-ambient sample stage.

**In-situ Neutron powder diffraction measurements.** In-situ temperature-dependent neutron powder diffraction data are collected using the time-of-flight Diffractometer GPPD (General Purpose Powder Diffractometer) at the China Spallation Neutron Source (CSNS) in Dongguan, China(*49, 50*).

**Processing of diffraction data for X-ray and Neutron.** The Rietveld refinement method is utilized to determine the crystal and magnetic structure using the General Structure Analysis System (GSAS) program(*51, 52*).

**Transmission electron microscope measurements.** The high-angle annular dark-field (HAADF) image and element distribution maps are obtained using a Tecnai G2 F20 S-TWIN transmission electron microscope (FEI, America).

**Scanning Transmission Electron Microscope.** The high-resolution scanning transmission electron microscope (STEM) image is obtained by spherical aberration-corrected scanning transmission electron microscope (Cs-STEM, Spectra 300, Thermo Fisher Scientific Inc., USA)

**Magnetic analysis.** A Quantum Design Physical Property Measurement System (PPMS, Quantum Design,

USA) is used to characterize the macroscopic magnetic properties in the magnetic field range from 8 T to -8 T. The temperature-dependent magnetization is measured during the warming and cooling processes under an applied magnetic field of 500 Oe.

# 6 Data availability

The raw/processed data required to reproduce these findings can be provided by the corresponding author upon reasonable request.

# 7 Code availability

The MATLAB code required to screen method of HEAPs can be provided by the corresponding author upon reasonable request.

# 9    Acknowledgments

This work was financially supported by the National Natural Science Foundation of China (NSFC) (Nos. 51972013, 52272264, 12074022 and 52371190), National Key R&D Program of China No. 2022YFA1402600, Sino-German Mobility Programme Project No. M-0273.

# 10 Author contributions

# 11 Competing interests

The authors declare no competing interests.

# 12 Additional information

Supplementary information

Supplementary Materials

# High-Entropy Enhanced Negative Thermal Expansion Perfomance in Antiperovkites

Xiuliang Yuan[a], Bing Wang[a], Ying Sun[a,*], Huaiming Guo[a], Kewen Shi[b], Sihao Deng[c,d,†], Lunhua He[c,e,f], Huiqing Lu[g], Hong Zhang[h,i,‡], Shengdi Xu[a], Yi Du[a], Weichang Hao[a], Shengqi Chu[d], Zhijie Ma[a], Shihai An[a], Jin Cui[b], Dongmei Hu[a], Huiming Han[a], Cong Wang[b,§]

a     *School of Physics, Beihang University, Beijing 100191, China*
b     *School of Integrated Circuit Science and Engineering, Beihang University, Beijing 100191, China*
c     *Spallation Neutron Source Science Center, Dongguan 523803, China*
d     *Institute of High Energy Physics, Chinese Academy of Sciences, Beijing 100049, China*
e     *Beijing National Laboratory for Condensed Matter Physics, Institute of Physics, Chinese Academic of Sciences, Beijing 100190, China.*
f     *Songshan Lake Materials Laboratory, Dongguan, Guangdong 523808, China*
g     *Key Laboratory for Photonic and Electronic Bandgap Materials, Ministry of Education, School of Physics and Electronic Engineering, Harbin Normal University, Harbin 150025, China*
h     *School of Materials and Energy, Electron Microscopy Centre, Yunnan University, Kunming 650500, P.R. China*
i     *School of Materials and Energy, Electron Microscopy Centre, Lanzhou University, Lanzhou 730000, P. R. China*

# 1    Predicted HEAP systems

Table S 1 Table 1 the predicted HEAPs by empirical rules based on the checkerboard diagram of solubility. The HEAPs in the column "Verified" have been verified experimentally and the HEAPs in the column "To be verified"

*   E-mail address: sunying@buaa.edu.cn

†   E-mail address: dengsh@ihep.ac.cn

‡   E-mail address: hongzhang@lzu.edu.cn

§   E-mail address: congwang@buaa.edu.cn

have not been verified yet.

| Predicted HEAPs | | | | |
|---|---|---|---|---|
| CuZnGaGeMn | MnCoNiCuGa | CoNiCuZnGa | MnNiCuZnAg | MnCuGaGeSn |
| CuZnGaGeSn | MnCoNiCuZn | CoNiCuZnAg | MnNiCuGaAg | MnCuGaAgSn |
| MnFeCoNiGa | MnCoNiCuAg | CoNiCuGaAg | MnNiCuGaFe | MnZnGaGeSn |
| MnFeCoNiCu | MnCoNiZnGa | MnCoCuZnGa | MnCuZnGaAg | MnZnGaAgSn |
| MnCoCuZnAg | MnCoNiZnAg | MnCoCuGaAg | MnCuZnGaSn | CoCuZnGaAg |
| MnNiZnGaAg | MnCoNiGaAg | MnCoCuGaFe | MnCuZnGeSn | NiCuZnGaAg |
|  | CoNiCuGaFe | MnCoZnGaAg | MnCuZnAgIn | CuZnGaAgSn |
|  | CoNiZnGaAg | MnNiCuZnGa | MnCuZnAgSn |  |

## 2 Room temperature X-ray diffraction of synthesized HEAP

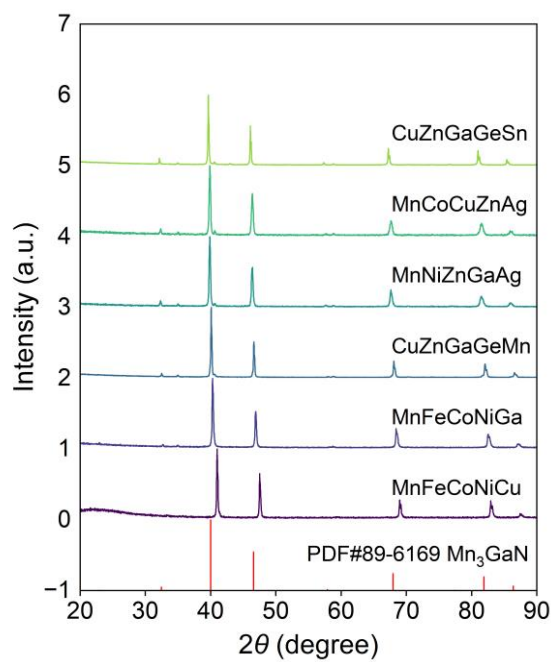

Figure S 1 X-ray diffraction patterns at room temperature for HEAPs: $Mn_3Mn_{0.2}Fe_{0.2}Co_{0.2}Ni_{0.2}Cu_{0.2}N$,

$Mn_3Mn_{0.2}Fe_{0.2}Co_{0.2}Ni_{0.2}Ga_{0.2}N$, $Mn_3Cu_{0.2}Zn_{0.2}Ga_{0.2}Ge_{0.2}Mn_{0.2}N$, $Mn_3Mn_{0.2}Ni_{0.2}Zn_{0.2}Ga_{0.2}Ag_{0.2}N$,

$Mn_3Mn_{0.2}Co_{0.2}Cu_{0.2}Zn_{0.2}Ag_{0.2}N$, $Mn_3Cu_{0.2}Zn_{0.2}Ga_{0.2}Ge_{0.2}Sn_{0.2}N$.

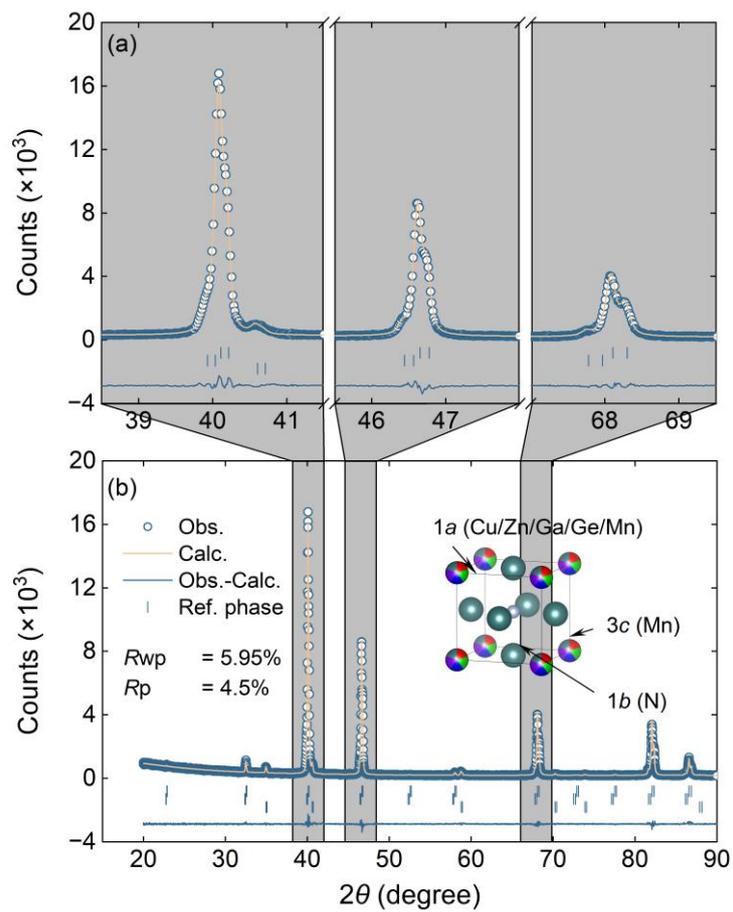

Figure S 2 the enlarged Rietveld refined XRD patterns of (b). (b) the result of Rietveld refinement of XRD patterns of $Mn_3Cu_{0.2}Zn_{0.2}Ga_{0.2}Ge_{0.2}Mn_{0.2}N$ at room temperature.

## 3 Synchrotron radiation X-ray diffraction

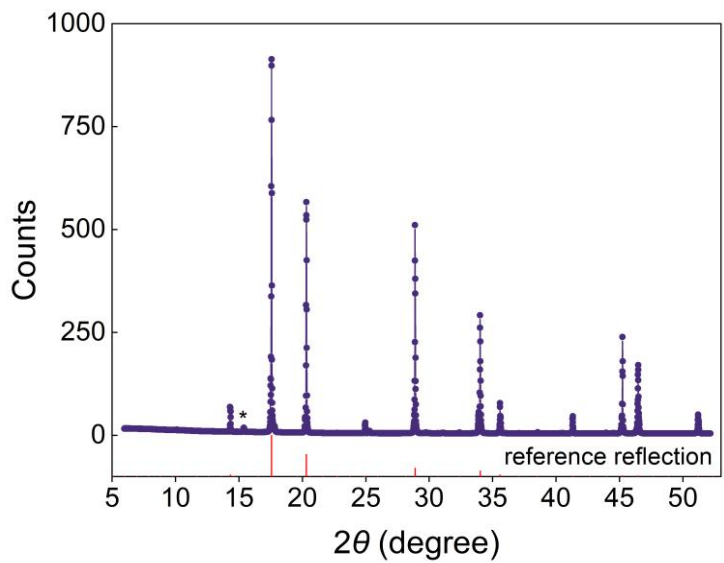

Figure S 3 Synchrotron radiation X-ray diffraction pattern of $Mn_3Cu_{0.2}Zn_{0.2}Ga_{0.2}Ge_{0.2}Mn_{0.2}N$ at room temperature.

## 4 TEM

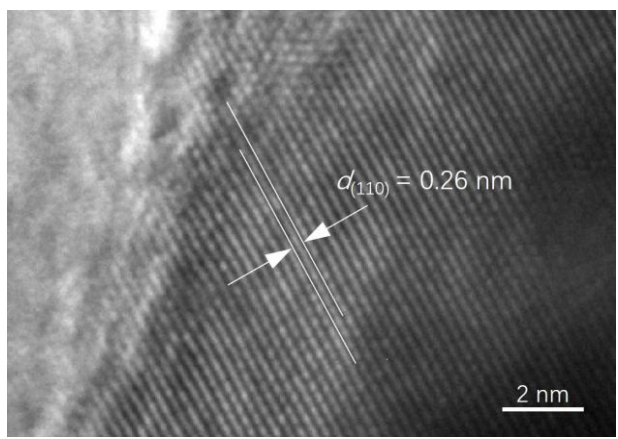

Figure S 4 TEM image of of $Mn_3Cu_{0.2}Zn_{0.2}Ga_{0.2}Ge_{0.2}Mn_{0.2}N$

# 5  EXAFS

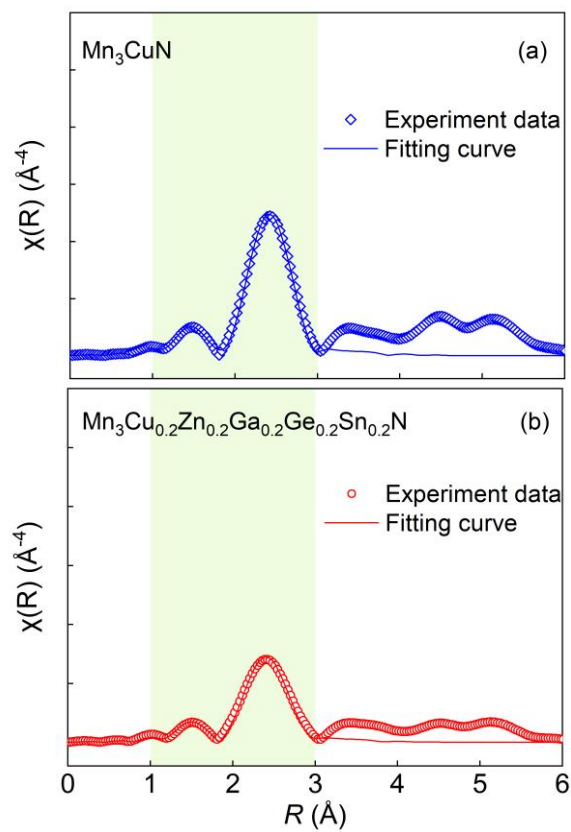

Figure S 5 EXAFS data for (a) reference $Mn_3CuN$ and (b) HEAP $Mn_3Cu_{0.2}Zn_{0.2}Ga_{0.2}Ge_{0.2}Sn_{0.2}N$.

Table S 2 analysis result of EXAFS data in Figure S 5. The error bar of interatomic distance for HEAP is significantly larger than the reference sample $Mn_3CuN$.

| Sample | Path | $N$ (fixed) | $R$ (Å) | $\sigma^2$ (Å$^2$) | $E_0$ (eV) | R factor |
|---|---|---|---|---|---|---|
|  | Mn-N | 2 | 1.947 (0.006) | 0.0031 (0.0022) | 9.9 (1.1) |  |
| $Mn_3CuN$ | Mn-Mn | 8 | 2.753 (0.009) | 0.0079 (0.0039) | 8.4 (1.1) | 0.003 |
|  | Mn-Cu | 4 | 2.753 (0.009) | 0.0082 (0.0061) | 8.4 (1.1) |  |
| $Mn_3Cu_{0.2}Z$ | Mn-N | 2 | 1.961 (0.019) | 0.000 (0.0033) | 9.9 (1.1) |  |
| $n_{0.2}Ga_{0.2}G$ | Mn-Mn | 8 | 2.749 (0.014) | 0.0073 (0.0060) | 6.1 (1.8) | 0.008 |
| $e_{0.2}Sn_{0.2}N$ | Mn-$X$ | 4 | 2.749 (0.014) | 0.0066 (0.0061) | 6.1 (1.8) |  |

# 6   Temperature dependent X-ray diffraction

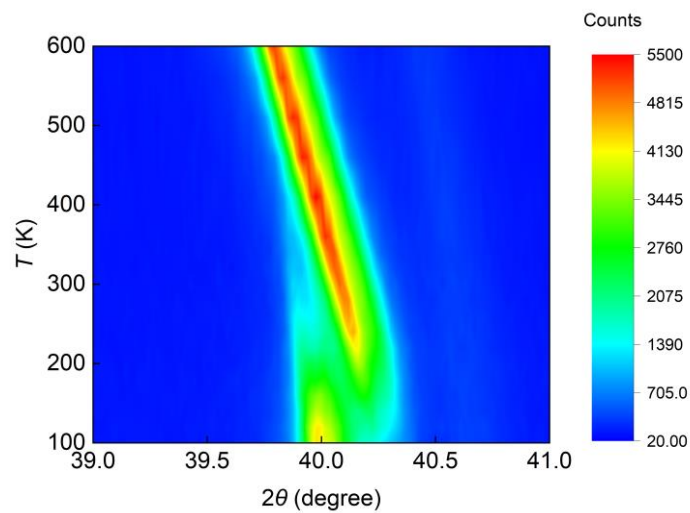

Figure S 6 temperature dependent XRD patterns of HEAP Mn$_3$Cu$_{0.2}$Zn$_{0.2}$Ga$_{0.2}$Ge$_{0.2}$Mn$_{0.2}$N.

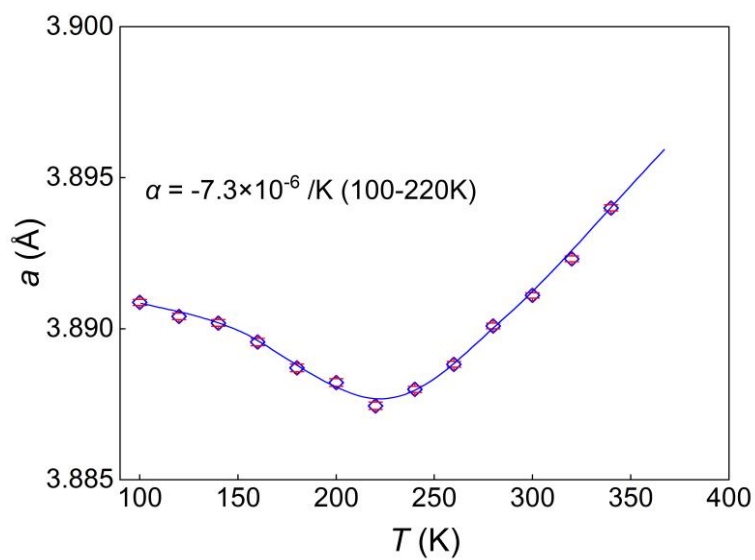

Figure S 7 temperature dependent lattice constant of HEAP Mn$_3$Cu$_{0.2}$Zn$_{0.2}$Ga$_{0.2}$Ge$_{0.2}$Mn$_{0.2}$N calculated using Figure S 6.

# 7  Analysis of STEM image

The raw STEM image shows indistinguishable contrast, making it difficult to determine the atom site of the disordered atom (1a site) and the Mn atom (3c site), see Figure S 8(a). However, after applying the Gaussian blur, a noticeable contrast becomes visible. Please refer to Figure S 8(b). The atom site of the disordered atom (1a site) and Mn atom (3c site) can be determined by comparing the scattering intensity of the atomic column of the disordered atom (1a site) with that of Mn (3c site). We add the crystal mask to the STEM image to illustrate the atomic positions (see Figure S 8(c)).

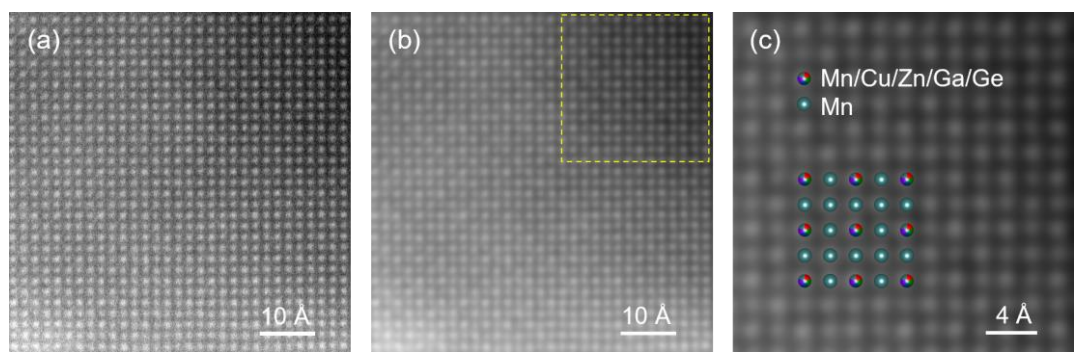

Figure S 8(a) raw STEM image of Mn3CuZnGaGeMnN along (001) direction. (b) STEM image after Gaussian blur. (c) The enlarged STEM image in (b) with the crystal structure mask.

We use the open-source StatSTEM software(*1*) (website: https://github.com/quantitativeTEM/StatSTEM) to conduct the subsequent analysis of lattice distortion. StatSTEM software provides a user-friendly method for quantifying atomic-resolution scanning transmission electron microscopy (STEM) images through the use of parametric model-based fitting. In this way, accurate and precise quantitative information can be extracted about the material being investigated. In StatSTEM, electron microscopy images are modeled by a superposition of Gaussian peaks that describe each atomic column. Hereby, the overlap of intensities from adjacent columns is taken into account. Structural parameters, such as column positions and scattering cross-sections, are determined and can be used for further analyses, such as atom counting, strain measurements, and creating 3D structural models.

The process of lattice distortion analysis is described in the following three steps.

- Preparation

    The initial fitting parameters for the next step of modeling fitting need to define the starting coordinates for the atomic column positions. In StatSTEM, the starting coordinates are determined by the local maxima in the image. Several filters, such as the average, disk, and Gaussian filters, are used to accurately determine the starting coordinates. The starting coordinates are plotted in Figure S 9(a) and (b).

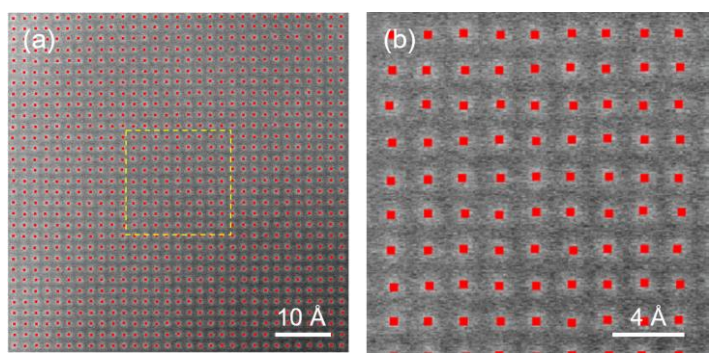

Figure S 9 (a) the starting coordinates for the atomic column positions and (b) its enlarged view. The red squares refer to the position of the local maxima.

- Modeling fitting

    The modeling fitting step aims to determine the fitted coordinates for the positions of the atomic columns. StatSTEM offers a standard procedure to model the STEM image using a superposition of Gaussian peaks. Notice that the column width of the Gaussian peaks is fixed to a consistent value for the same atom type. This approach is computationally less demanding compared to fitting all the peaks with varying widths, and it yields reliable results. To enhance computational speed, parallel computing with 4 CPU cores is utilized. The fitted

STEM image shows less noise compared to the raw data (see Figure S 10(a) and (b)).

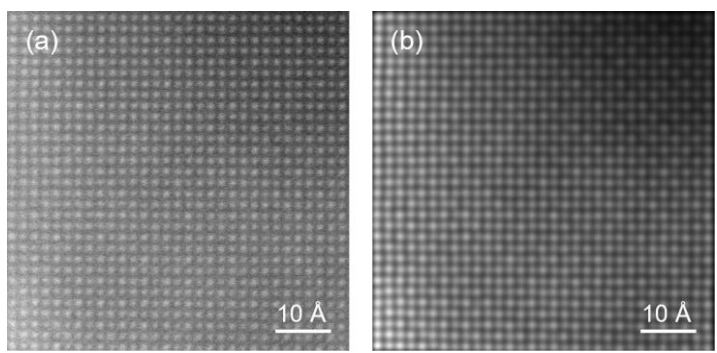

Figure S 10(a) raw STEM image. (b) the fitted STEM image.

The fitted coordinates for the atomic column positions are slightly different compared to the starting coordinates in Figure S 9(a) and (b). We illustrate the difference in an enlarged image (see Figure S 11(a) and (b)).

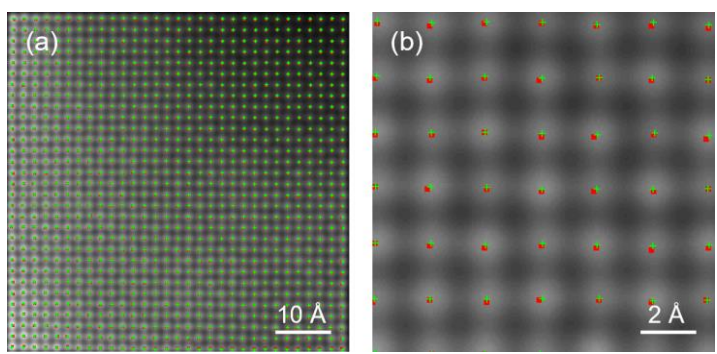

Figure S 11 (a) the comparison between starting coordinates and fitted coordinates for the atomic column positions and (b) its enlarged view. The red squares refer to the position of starting coordinates and the green cross refer to the position of fitted coordinates.

- Analysis

The analysis of lattice distortion requires calculating the deviation of the actual position (the fitted coordinates)

from the reference position of the atom column. The reference position is determined by the definition of the projected unit cell. The lattice parameters in the projected unit cell should closely match the experimental values. For the present HEAP, a suitable value is 4.078 Å. We plot the reference and actual positions of the atom column in Figure S 12(a) and (b).

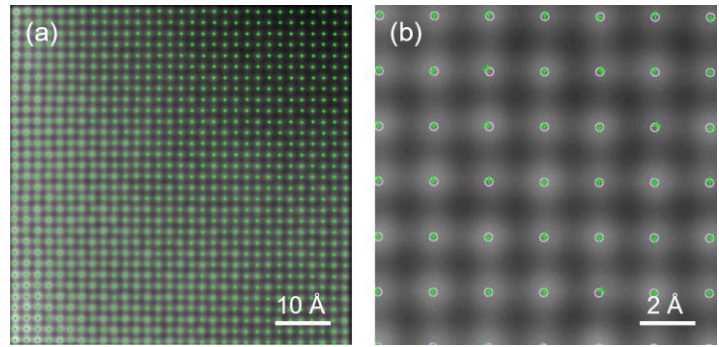

Figure S 12 the comparison between reference coordinates and actual (fitted) coordinates for the atomic column positions and (b) its enlarged view. The pink circle refers to the position of reference coordinates and the green cross refer to the position of actual (fitted) coordinates.

The deviation between the actual position and reference position is used to draw the lattice distortion diagram including magnitude and direction. By calculating the atomic spacing, the spacing fluctuation can be quantized for Mn (at $3c$ site) and disordered atoms (at $1a$ site). In Figure S 13, we give the distribution of displacement magnitude for HEAP and reference MnO.

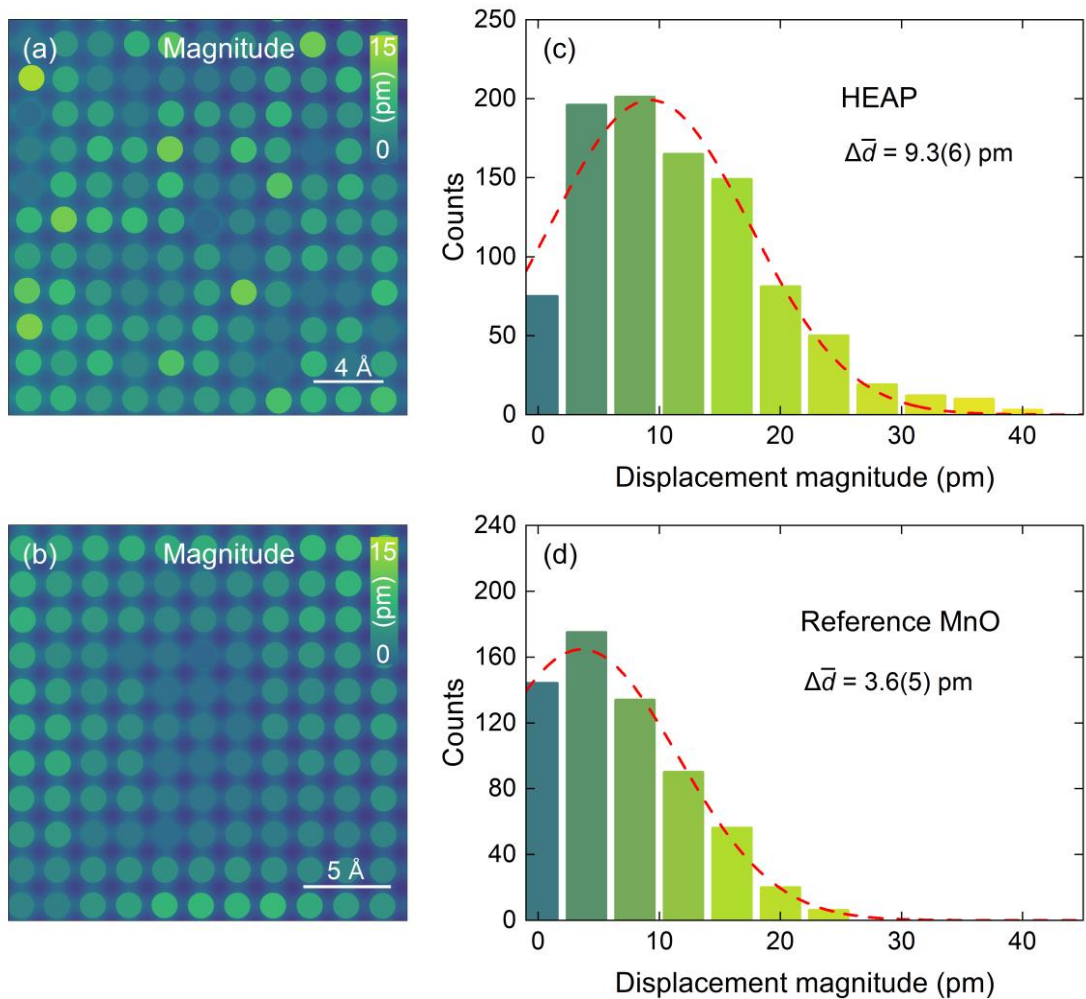

Figure S 13 (a) lattice distortion quantized in magnitude and (b) direction. (c) distribution of displacement magnitude for HEAP and (d) reference MnO.